# Geometric effect and gauge field in nonequilibrium quantum thermostatistics

Sumiyoshi Abe* and Satoshi Kaneko**

*Institute of Physics*, *University of Tsukuba*, *Ibaraki 305-8571*, *Japan*

**Abstract**   The concept of work is studied in quantum thermostatistics of a system surrounded by an environment and driven by an external force. It is found that there emerges the gauge theoretical structure in a nonequilibrium process, the field of which is referred to as the *work gauge field*. The thermodynamic work as the flux of the work gauge field is considered for a cyclic process in the space of the external-force parameters. As an example, the system of a spin-1/2 interacting with an external magnetic field is analyzed. This geometric effect may be observed, for example, in an NMR experiment and can be applied to the problem of cooling/heating of a small system.

Recent developments in dynamically manipulating nanosystems have been drawing fresh and general interest in thermostatistics in a small scale. From the physics viewpoint, they seem to require deeper understandings of nonequilibrium thermostatistics. In this direction, some intriguing discussions have been made in the literature. Among others, an attempt of Jarzynski [1] to connect nonequilibrium works with equilibrium free energies has been attracting much attention both theoretically [2] and experimentally [3]. It leads to necessity of reexamining the basic thermodynamic quantities in the nonequilibrium regime.

In this paper, we study the concept of work in nonequilibrium quantum thermostatistics. We find that there naturally emerges the gauge structure in the theory, the field of which is referred to as the *work gauge field*. In particular, we consider a cyclic process in the space of the parameters, which describe the external driving of the system. Then, the work turns out to be given by the flux of the work gauge field. We examine this general result for an example of a spin-1/2 interacting with an external magnetic field. We also make comments on a possibility of observing such a geometric effect, for example, in an NMR experiment and its relevance to the problem of cooling/heating of a small system.

Let us start our discussion with the quantum mechanical formulation of the first law of thermodynamics. Consider a driven quantum system in contact with an environment. Its state is represented by a density matrix $\rho$, which is a Hermitian, traceclass, positive semidefinite operator. The internal energy is given by the expectation value of the system Hamiltonian $H$ with respect to $\rho$: $U = \text{Tr}(\rho H)$. Its infinitesimal change is given by

$$dU = \text{Tr}(d\rho\, H) + \text{Tr}(\rho\, dH). \tag{1}$$

Then, the first law of thermodynamics, $d'Q = dU + d'W$, is realized if the following identifications are made for the work and the quantity of heat:

$$d'W = -\text{Tr}(\rho\, dH), \tag{2}$$

$$d'Q = \text{Tr}(d\rho\, H). \tag{3}$$

The external driving is described by the dependency of the Hamiltonian on a set of the parameters

$$H = H(\lambda^1, \lambda^2, \cdots, \lambda^n). \tag{4}$$

Note that, in contrast to Ref. [1], here we are considering the multiple parameters. Accordingly, the work in Eq. (2) is written as

$$d'W = \sum_{\mu=1}^{n} a_\mu\, d\lambda^\mu, \tag{5}$$

where

$$a_\mu = -\text{Tr}(\rho\, \partial_\mu H) \qquad (6)$$

with the notation, $\partial_\mu \equiv \partial/\partial\lambda^\mu$. As we shall see below, the quantity in Eq. (6) is a gauge field potential, which is referred to here as the work gauge field. It may be analogous to the gauge field in the discussion of the Berry phase [4] in the sense that it is induced by the external parameters.

Now, for a process along a closed curve $C$ in the parameter space, the work is written as

$$W_C = \sum_{\mu=1}^{n} \oint_C a_\mu\, d\lambda^\mu = \frac{1}{2} \sum_{\mu,\nu=1}^{n} \iint_S f_{\mu\nu}\, d\lambda^\mu \wedge d\lambda^\nu. \qquad (7)$$

where $f_{\mu\nu} \equiv \partial_\mu a_\nu - \partial_\nu a_\mu$ and $C = \partial S$ (i.e., $C$ is the boundary of the surface $S$ in the parameter space).

Clearly, the "field strength", $f_{\mu\nu}$, vanishes (and thus $W_C = 0$) if the density matrix is a function only of the Hamiltonian. Therefore, for example, the equilibrium state represented by the canonical density matrix

$$\rho_{eq} = \frac{1}{Z(\beta)} e^{-\beta H} \qquad (8)$$

with the partition function, $Z(\beta) = \text{Tr}\, e^{-\beta H}$, does not yield the nonvanishing field strength. In fact, for $\rho_{eq}$, the work gauge field becomes pure gauge, $a_\mu = -\partial_\mu F$, where $F = -\beta^{-1} \ln Z(\beta)$ is the free energy. This is nothing but a familiar relation that the work done is given by the free energy difference in an isothermal process.

The above discussion tells us that the associated gauge transformation of the density matrix is

$$\rho \rightarrow \rho + \tilde{\rho}, \tag{9}$$

where $\tilde{\rho}$ is a traceless matrix depending only on the Hamiltonian and does not violate the positive semidefiniteness of the transformed density matrix. Under Eq. (9), the work gauge field changes as follows:

$$a_\mu \rightarrow a_\mu + \partial_\mu \Lambda, \tag{10}$$

where $\partial_\mu \Lambda \equiv -\text{Tr}[\tilde{\rho}(H) \partial_\mu H]$ (such $\Lambda$ exists since $\tilde{\rho}$ is a function only of the Hamiltonian). $a_\mu$ is, therefore, an Abelian gauge field. The work in Eq. (7) is a gauge-invariant quantity.

Thus, the work gauge field can give rise to a nonvanishing field strength if the state is out of equilibrium, in general. To realize such a nontrivial work gauge field, we employ a quantum operation [5] on the equilibrium canonical density matrix in Eq. (8). That is, $\rho_{eq}$ is linearly mapped to a nonequilibrium state $\rho$ as follows:

$$\rho_{eq} \rightarrow \rho = \Phi(\rho_{eq}). \tag{11}$$

In particular, we represent the map by using the positive operator-valued measure (POVM) [5]

$$\rho = \Phi(\rho_{eq}) = \sum_k V_k \, \rho_{eq} \, V_k^\dagger. \tag{12}$$

It satisfies the trace-preserving condition, $\sum_k V_k^\dagger V_k = I$, where $I$ stands for the identity operator and $V_k$'s may depend on the set of the parameters $\{\lambda_i\}_{i=1,2,\cdots,n}$. To obtain a nonequilibrium state, clearly $V_k$'s cannot be functions only of the Hamiltonian. Unlike in the equilibrium state $\rho_{eq}$, $\beta$ does not have the meaning as the inverse temperature in the nonequilibrium state $\rho$, in general. Substituting Eq. (12) into Eq. (6), we obtain the nonvanishing work, $W_C$.

We wish to notice two point, here. One is that there are a variety of nonequilibrium states, and the present construction is, though quite general, just one example. However, the above one is useful since it is in reference to the equilibrium state. The other is that, due to the trace-preserving condition, a monopole-type singularity does not appear in the configuration of the work gauge field.

Finally, let us consider an example of a single spin-1/2 interacting with a magnetic field $\mathbf{B} = (B_x, B_y, B_z)$, which plays a role of the external parameters. The system Hamiltonian reads

$$H = \kappa \sigma \cdot \mathbf{B}. \tag{13}$$

Here $\sigma$'s are the Pauli matrices and $\kappa$ is a constant involving the magnetic moment. The Planck constant as a multiplicative factor is set equal to unity for the sake of simplicity. The equilibrium canonical density matrix is given by

$$\rho_{eq} = \frac{1}{2}\left[I - \frac{\sigma \cdot \mathbf{B}}{B}\tanh(\beta \kappa B)\right], \tag{14}$$

where $B = |\mathbf{B}|$.

To map the equilibrium state in Eq. (14) to a nonequilibrium one, we consider the following most general POVM:

$$V_k = c_k I + \mathbf{X}_k \cdot \sigma, \tag{15}$$

where $c_k$'s and $\mathbf{X}_k$'s are complex numbers and complex vectors generically dependent on the magnetic field, respectively. The trace-preserving condition leads to

$$\sum_k \left(|c_k|^2 + \mathbf{X}_k^* \cdot \mathbf{X}_k\right) = 1, \tag{16}$$

$$\sum_k \left(c_k^* \mathbf{X}_k + c_k \mathbf{X}_k^* + i\mathbf{X}_k^* \times \mathbf{X}_k\right) = \mathbf{0}. \tag{17}$$

Using the nonequilibrium density matrix $\rho = \sum_k V_k \rho_{eq} V_k^\dagger$ with Eqs. (14) and (15), we find that the work gauge field is given by

$$\mathbf{a}(\mathbf{B}) = -\kappa \sum_k \left( c_k \mathbf{X}_k^* + c_k^* \mathbf{X}_k + i \mathbf{X}_k \times \mathbf{X}_k^* \right)$$

$$+ \kappa \frac{\tanh(\beta \kappa B)}{B} \sum_k \left[ |c_k|^2 \mathbf{B} + i c_k^* \mathbf{X}_k \times \mathbf{B} + i c_k \mathbf{B} \times \mathbf{X}_k^* \right.$$

$$\left. + (\mathbf{B} \cdot \mathbf{X}_k) \mathbf{X}_k^* + (\mathbf{B} \cdot \mathbf{X}_k^*) \mathbf{X}_k - (\mathbf{X}_k^* \cdot \mathbf{X}_k) \mathbf{B} \right]. \tag{18}$$

This field is drastically simplified if the following choice is made:

$$\mathbf{X}_k = (0, 0, g_k), \tag{19}$$

where $g_k$'s, as well as $c_k$'s, are real constants. In this case, the condition in Eq. (16) and (17) become

$$\sum_k \left( c_k^2 + g_k^2 \right) = 1, \tag{20}$$

$$\sum_k c_k g_k = 0, \tag{21}$$

respectively. Now, performing a straightforward calculation, we obtain the work gauge field of the following form:

$$\mathbf{a}(\mathbf{B}) = f(B)(\alpha B_x, \alpha B_y, B_z), \tag{22}$$

where

$$f(B) = \kappa \frac{\tanh(\beta \kappa B)}{B}, \tag{23}$$

$$\alpha = \sum_k \left( c_k^2 - g_k^2 \right). \tag{24}$$

Taking the rotation of this field, $\mathbf{b}(\mathbf{B}) = \nabla_B \times \mathbf{a}(\mathbf{B})$ with the notation $\nabla_B \equiv \partial/\partial \mathbf{B}$, we have the following field strength:

$$\mathbf{b}(\mathbf{B}) = (\alpha - 1) \frac{B_z \sqrt{B_x^2 + B_y^2}}{B} \frac{d f(B)}{d B} \mathbf{e}_\phi, \tag{25}$$

where $\mathbf{e}_\phi$ is the unit vector in the direction of rotation around the $B_z$-axis, that is,

$$\mathbf{e}_\phi = \left( -\frac{B_y}{\sqrt{B_x^2 + B_y^2}}, \frac{B_x}{\sqrt{B_x^2 + B_y^2}}, 0 \right). \tag{26}$$

For a process along a closed curve $C$ in the $\mathbf{B}$-space, the work is given in terms of the flux of $\mathbf{b}$ traversing the surface $S$ surrounded by $C$. In other words, given the work gauge field strength in Eq. (23), the work is determined sorely by the geometric configuration of $C$.

In a particular case when all $c_k$'s vanish, $\alpha = -1$ and the quantum operation becomes isoentropic: $\Phi(\rho_{eq}) = \sigma_z \rho_{eq} \sigma_z$.

In conclusion, we have studied the concept of work in quantum thermostatistics of a system driven by an external force. We have found that the gauge theoretical structure naturally emerges in a nonequilibrium process, which is generated by using the positive operator-valued measure (POVM). We have described the thermodynamic work as the flux of the gauge field termed the work gauge field and have considered it for a cyclic process in the space of the external-force parameters. The example of a spin-1/2 interacting with an external magnetic field has been explicitly analyzed. It is our hope that such an effect can be observed, for example, in an NMR experiment.

Since the external parameters and the POVM operators are experimentally controllable, one can, for example, reverse the direction of the cyclic process $C$. Under such a reversal, $W_C$ changes its sign. Therefore, the present study can be applied to the problem of cooling/heating of a small system, which is of physical relevance to nanoscience.

———————————————